\documentclass[lettersize,journal]{IEEEtran}
\usepackage{amsmath,amsfonts}
\usepackage{algorithmic}
\usepackage{algorithm}
\usepackage{array}
\usepackage[caption=false,font=normalsize,labelfont=sf,textfont=sf]{subfig}
\usepackage{textcomp}
\usepackage{stfloats}
\usepackage{url}
\usepackage{verbatim}
\usepackage{graphicx}
\usepackage{booktabs}%
\usepackage{multirow}%
\usepackage{tabularx}

\usepackage{adjustbox}

\usepackage[
  style=ieee,
  backend=biber,
  abbreviate=true,
  isbn=false,
  doi=false,
  url=false,
  eprint=false,
]{biblatex}

\usepackage{balance}

\usepackage{etoolbox}
\newcommand{\tablefontsize}{\footnotesize}
\AtBeginEnvironment{table}{\tablefontsize}
\AtBeginEnvironment{longtable}{\tablefontsize}

\renewbibmacro*{journal}{%
  \iffieldundef{shortjournal}
    {\printfield{journaltitle}}
    {\printfield{shortjournal}}%
}

\renewbibmacro*{journal+issuetitle}{%
  \usebibmacro{journal}%
  \setunit*{\addcomma\space}%
  \iffieldundef{series}
    {}
    {\newunit
     \printfield{series}%
     \setunit{\addspace}}%
  \usebibmacro{volume+number+eid}%
  \setunit{\addspace}%
  \usebibmacro{issue+date}%
  \setunit{\addcolon\space}%
  \usebibmacro{issue}%
  \newunit}

\DeclareFieldFormat{pages}{#1}

\renewbibmacro*{volume+number+eid}{%
  \printfield{volume}%
  \setunit*{\addnbspace}%
  \printfield{number}%
  \setunit{\addcomma\space}%
  \printfield{eid}%
}

\DeclareFieldFormat[online]{title}{\mkbibquote{#1}}

\renewbibmacro*{volume+number+eid}{%
  \printfield{volume}%
  \setunit*{\addnbspace}%
  \printfield{number}%
  \setunit{\addcomma\space}%
  \printfield{eid}%
}

\usepackage{longtable}

\begin{document}

\title{The Shadow of Fraud: The Emerging Danger of AI-powered Social Engineering and its \\ Possible Cure}

\author{
Jingru Yu$^{1}$,Yi Yu$^{1}$,Xuhong Wang$^{1}$,Yilun Lin$^{1,*}$, \textit{Member, IEEE}, Manzhi Yang$^{2,3}$,Yu Qiao$^{1}$,\\ Fei-Yue Wang$^{2,4}$, \textit{Fellow, IEEE} 
\thanks{This work is supported by Shanghai Artificial Intelligence Lab}
\thanks{* for the corresponding author}
\thanks{$^{1}$Jingru Yu (yujingru@pjlab.org.cn), Yi Yu (yuyi@pjlab.org.cn), Xuhong Wang (wangxuhong@pjlab.org.cn), Yilun Lin (linyilun@pjlab.org.cn) and Yu Qiao (qiaoyu@pjlab.org.cn) are with Shanghai Artificial Intelligence Laboratory, Shanghai 200240, China.}
\thanks{$^{2}$ Manzhi Yang (yangmanzhi@eversec.cn) and Fei-Yue Wang (feiyue.wang@ia.ac.cn) are with the Macau Institute of Systems Engineering, Macau University of Science and Technology, Macau, China}
\thanks{$^{3}$ Manzhi Yang is also with  Eversec Technology Company Ltd., Beijing 100080, China}
\thanks{$^{4}$ Fei-Yue Wang is also with the State Key Laboratory for Management and Control of Complex Systems, Institute of Automation, Chinese Academy of Sciences, Beijing 100190, China}
\thanks{Manuscript received June 10, 2024.}
}

\markboth{Journal of \LaTeX\ Class Files,~Vol.~14, No.~8, August~2021}%
{Shell \MakeLowercase{\textit{et al.}}: A Sample Article Using IEEEtran.cls for IEEE Journals}

\maketitle

\begin{abstract}
Social engineering (SE) attacks remain a significant threat to both individuals and organizations. 
The advancement of Artificial Intelligence (AI), including diffusion models and large language models (LLMs), has potentially intensified these threats by enabling more personalized and convincing attacks.
This survey paper categorizes SE attack mechanisms, analyzes their evolution, and explores methods for measuring these threats. It highlights the challenges in raising awareness about the risks of AI-enhanced SE attacks and offers insights into developing proactive and adaptable defense strategies.
Additionally, we introduce a categorization of the evolving nature of AI-powered social engineering attacks into "3E phases": Enlarging, wherein the magnitude of attacks expands through the leverage of digital media; Enriching, introducing novel attack vectors and techniques; and Emerging, signifying the advent of novel threats and methods. 
Moreover, we emphasize the necessity for a robust framework to assess the risk of AI-powered SE attacks. By identifying and addressing gaps in existing research, we aim to guide future studies and encourage the development of more effective defenses against the growing threat of AI-powered social engineering.
\end{abstract}


\section{Introduction}
The rapid advancement of artificial intelligence has revolutionized various aspects of our lives, including how we communicate, interact, and exchange information. However, this technological progress has also given rise to a new and insidious threat: AI-powered social engineering   attacks~\cite{king2020artificial}. 
The 2023 Internet Crime Report from the Federal Bureau of Investigation (FBI) revealed that losses from tech support fraud, one form of AI-powered social engineering attacks, totaled \$924.5 million in 2023, representing the third most costly crime category according to Internet Crime Complaint Center (IC3) statistics~\cite{crime}. In addition, over 21,000 complaints of business email compromise, another type of social engineering threat, resulted in \$2.9 billion in damages~\cite{crime}.
Notably, preliminary research has demonstrated AI systems' latent abilities to conduct social engineering clandestinely. 
For instance, DeepLocker~\cite{stoecklin2018deeplocker} has demonstrated the potential for using AI to carry out highly targeted and covert SE attacks within seemingly harmless applications. Furthermore, automated social engineering (ASE)~\cite{king2020artificial}has exacerbated this issue by using tools such as bots to automatically execute social engineering attacks. As AI systems become more sophisticated, they possess the ability to manipulate human behavior and exploit our inherent biases and vulnerabilities, posing a significant risk.

\begin{figure} [htbp]
    \centering
    \includegraphics[width=0.9\linewidth]{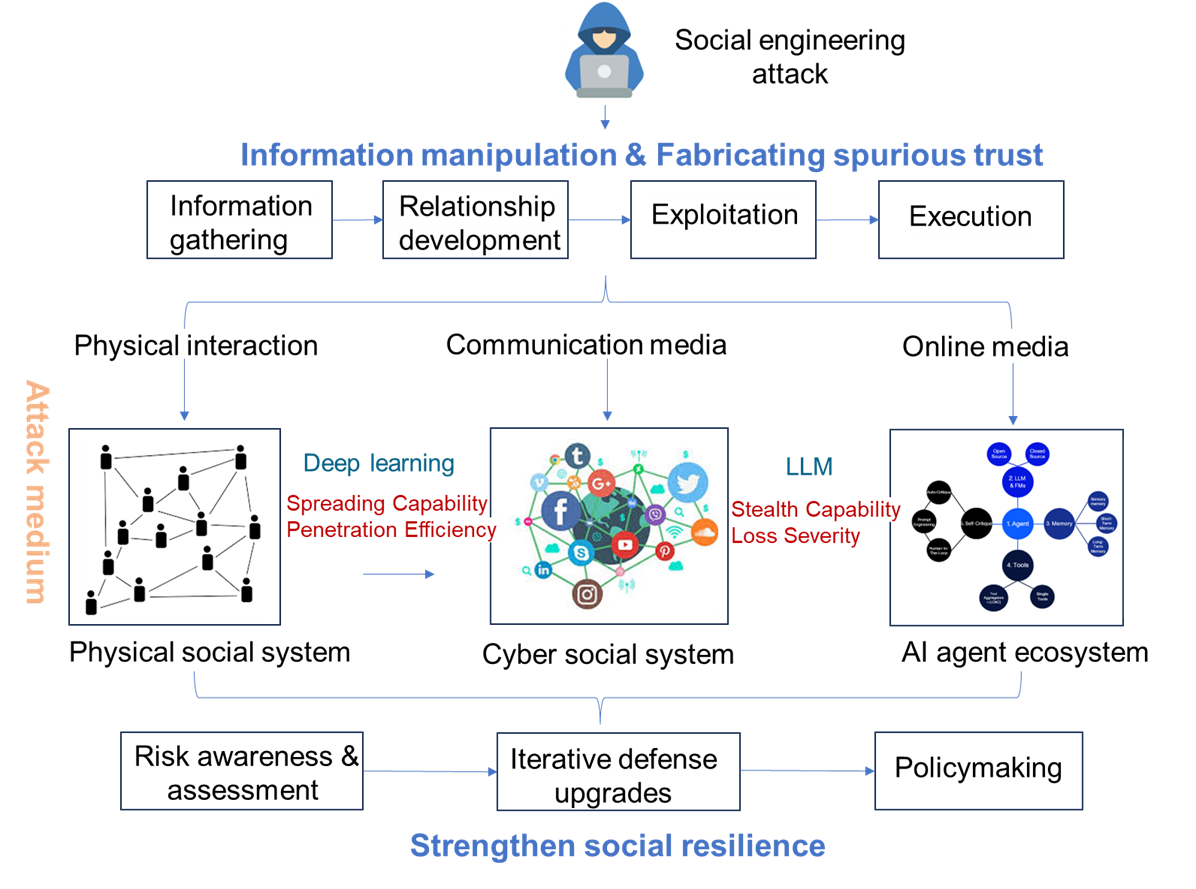}
    \caption{Advances in AI-powered social engineering attack in the context of evolving social systems.}
    \label{intro}
\end{figure}

Social engineering attacks is about exploiting human psychological and behavioral characteristics to bypass technical security measures. Generally, such attacks follow a four-step process: information gathering, relationship building, exploitation, and execution~\cite{steinmetz2021performing}. 
As information transmission media and social organizations evolve, SE attack medium has shifted from physical to digital communication~\cite{wang2021social} and now agent-based platforms, as shown in Fig.~\ref{intro}.  
In early physical system, attackers relied on direct interaction, conferring limited outreach. And digitization catalyzed an Enlarging phase by automating traditional attack techniques at scale.
Advanced AI models such as generative adversarial networks(GAN) and Reinforcement Learning(RL) algorithms heralded an Enriching transition, empowering personalized deception through mechanisms like highly targeted phishing. These attacks can bypass traditional security measures and exploit the trust and goodwill of targeted individuals, luring them into divulging sensitive information or performing actions that compromise the integrity of their accounts or systems.
Moreover, noval attack forms have emerged along with the
emergence of intelligent agents.
Currently, hacking techniques become easily reached. Tools like WormGPT~\cite{firdhous2023wormgpt} and FraudGPT enable automated, vast-scale attacks via LLM, employed in cybercrimes including ransomware~\cite{firdhous2023wormgpt}.
Accordingly, the SE threat landscape undergoes progressive 3E transformations. The first phase involved an enlargement of scale through digitization. The second phase brought an enrichment of attack vectors using profiling. The third phase could see the emergence of qualitatively different deception modes. Such modes may involve sophisticated artificial intelligence technologies.

Several surveys have explored SE attacks and those enabled by AI~\cite{salahdine2019social, mashtalyar2021social, fuertes2022impact, salama2023social}. Salahdine et al.(2019)~\cite{salahdine2019social} presents an in-depth survey about social engineering attacks, existing detection methods, and countermeasure techniques. However, the intrinsic link between AI risks and SE attacks went unexamined. Kaloudi et al. (2020)~\cite{kaloudi2020ai} classified AI's malicious uses in cyber attacks but lacked case studies illustrating real-world impacts The review by Blauth et al. (2022)~\cite{blauth2022artificial} served as an introduction yet failed to fully capture the evolving landscape of AI-powered SE attacks, especially regarding LLM and associated emerging risks. Therefore, there exist research gaps in the following aspects: tracking and analyzing the rapidly changing landscape of AI-powered SE attacks, developing a robust framework for quantitatively understanding the AI risk, developing proactive and adaptive detection and defense strategies, and addressing the ethical and privacy implications of AI-powered SE attacks. Our paper aims to address the abovementioned gaps identified in the previous studies, and its contributions are as follows:

\begin{itemize}
\item\textbf{Consolidation of existing knowledge} This study consolidates existing AI-powered SE attack research, tracking their implementation mechanisms, and technical evolution from traditional SE attack era to LLM era. 
\item \textbf{Insight into AI risk awareness and quantification}  As the forms of SE attacks continue to evolve, traditional risk evaluation methods may no longer be applicable, requiring the development of risk quantification mechanisms to accurately measure their potential threats. We propose a risk quantification framework to support AI risk awareness and policy-making.
\item \textbf{Comprehensive taxonomy of defense challenges} This work goes beyond the scope of existing SE attack studies by offering a taxonomy for SE attack defense techniques. The taxonomy will guide future studies on developing more resilient defense techniques.
\end{itemize}

The rest of this paper is organized as follows. Section 2 outlines the literature review methodology, including search scope, strategy and key findings.
Section 3 characterizes SE attack evolution through three intersecting phases: the initial Enlarging of scope through online proliferation, its subsequent Enriching of attack vectors such as deepfake and social media bots, and potential Emergence of novel attack modes leveraging advanced LLM. Representative SE attack techniques are examined within each transformational stage to contextualize shifting methodologies. 
Section 4 briefly analyzes challenges, discusses future prospects, and outlines promising directions for mitigating emerging threats posed by increasingly sophisticated AI capabilities. Section 5concludes the work by summarizing the paper's contributions towards systematically organizing extant research and anticipating future directions.
\section{Systematic literature review methodology}\label{sec2}

\subsection{Survey Scope}

This survey aims to provide a comprehensive overview of the current state of research on the AI-powered social engineering attacks. As shown in Fig.~\ref{scope}, the scope of this review includes the following key aspects:

\begin{figure*}[ht]
\centering
\includegraphics[width=0.9\linewidth]{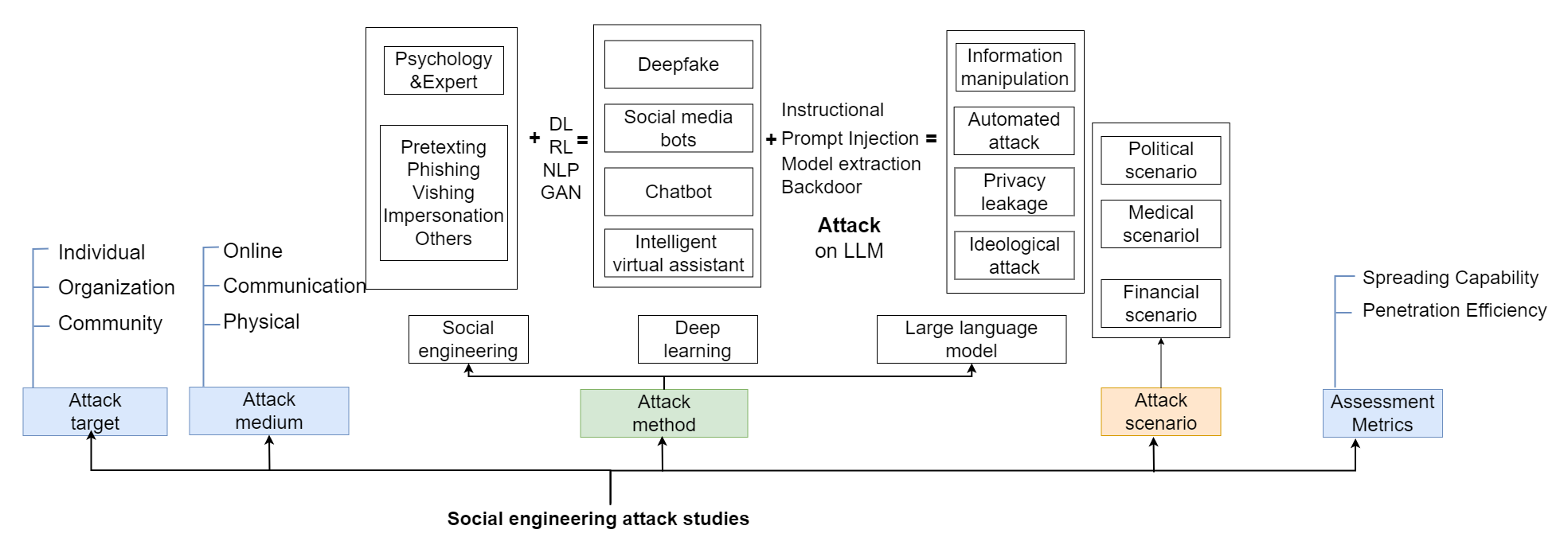}
\caption{Survey scope: establishing  understanding of SE attack evolution, analyzing connections to emerging AI risks and examining real-world case studies.}
\label{scope}  
\end{figure*}
\begin{itemize}
\item Establishing a comprehensive understanding of AI-powered SE attack based on the in-depth examination of the definition, targets, mediums, methods, and specific scenarios to support AI risk awareness and quantification. 

\item Analyzing the inherent connection between emerging AI-related risks and their influence on the workflow and impact of social engineering attacks.

\item Examining case studies that illustrate the real-world implications and consequences of AI-powered SE attacks. These case studies can shed light on the practical manifestations of this threat and facilitate a deeper understanding.

\item Identifying the key research challenges in tracking and analyzing the rapidly changing landscape of AI-powered SE attacks. 

\end{itemize}

Basic concepts related to SE attacks are described as the following:
\begin{itemize}
\item\textbf{Social engineering attack} Social engineering attacks are rapidly increasing in today’s networks and are weakening the cybersecurity chain. They aim at manipulating individuals and enterprises to divulge valuable and sensitive data in the interest of cyber criminals~\cite{salahdine2019social}.

\item \textbf{Attack targets} Attack target is the party to suffer a social engineering attack and bring about an attack consequence. 
\item \textbf{Attack medium} Attack medium is the entity for the social interaction to implement (through which the target is contacted), and the substance or channel through which attack methods are carried out~\cite{wang2021social}. 


\item \textbf{Attack method}
Attack method is the way, manner or means of carrying an attack out. Synonyms such as attack vector, attack technique and attack approach are used to convey the same meaning.
\end{itemize}

\subsection{Paper collection strategy}

A rigorous search strategy was adopted to comprehensively survey the literature on AI-powered social engineering attacks and associated defenses. The following electronic databases were queried:
IEEE Xplore, Google Scholar, ACM Digital Library, SpringerLink, arXiv, CVPR,  ScienceDirect(ELSEVIER) and Web of Science. Relevant literature from 2000 to present totals 473 papers.  The keywords we used are as follows:
\begin{itemize}
\item “social engineering”/“fake content”/“information manipulation”
\item “AI for social engineering”/“LLM for social engineering”
\item “social engineering detection”/ “fake content detection”/“social engineering defense challenges”
\end{itemize}

\subsection{Findings}

\begin{figure} [htbp]
    \centering
    \includegraphics[width=0.9\linewidth]{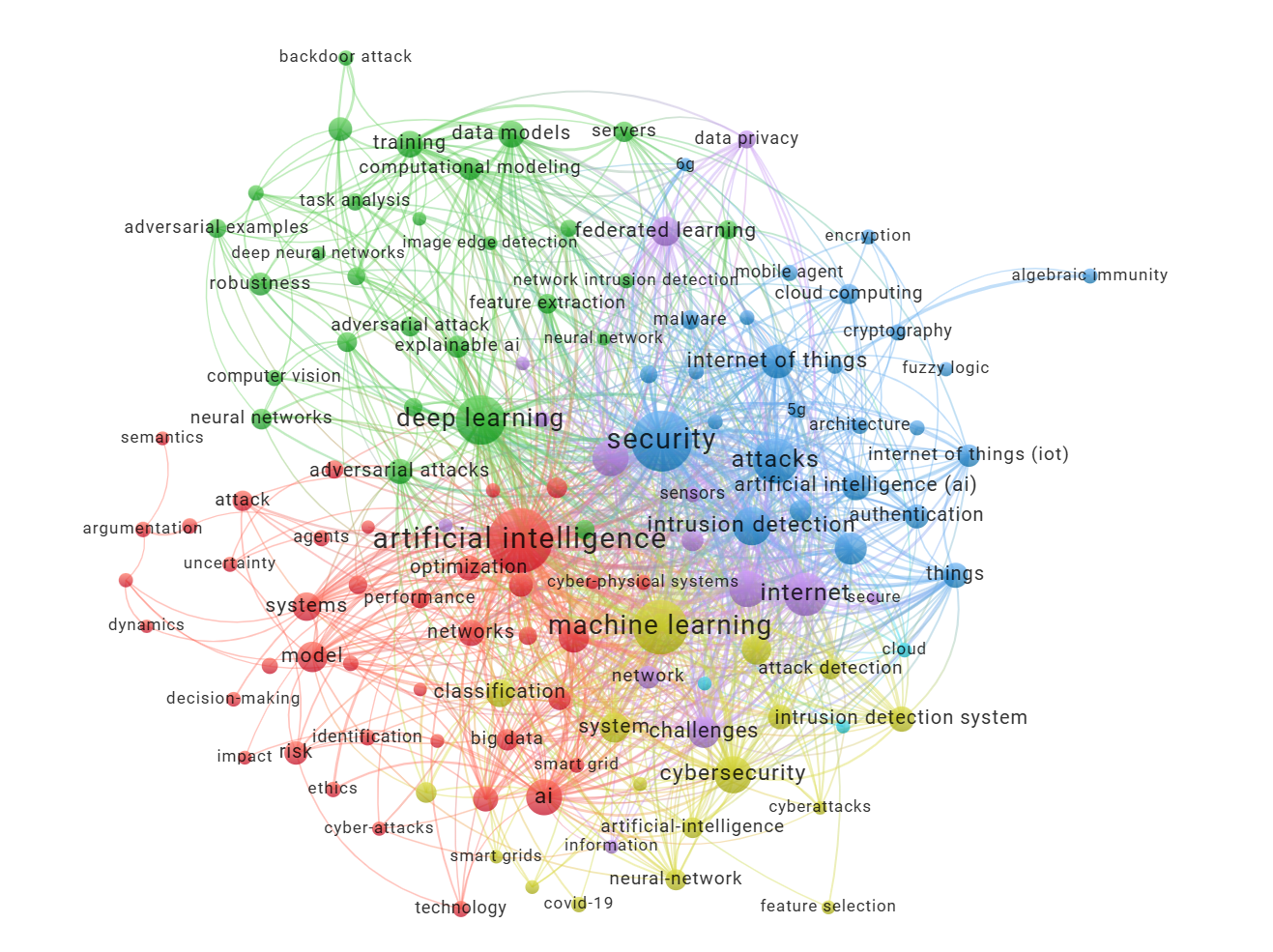}
    \caption{Keyword co-occurrence analysis in social engineering research.}
    \label{key}
\end{figure}

We perform a co-occurrence analysis of keywords in the literature using the
VOSviewer software~\cite{vaneck_software_2010}, as depicted in Fig.~\ref{key}. The keywords exhibit a clear temporal transition trend, which can be
categorized into as follows:
\begin{itemize}
\item 
Red terms focus on AI techniques (artificial intelligence, machine learning, big data) and application domains (decision making, internet, security, privacy) that are relevant to SE attacks. This reflects increasing exploration of advanced approaches to automate and enhance attack capabilities.
\item 
Blue terms involve core technological research areas (security, architecture, IoT, authentication) that are being investigated to defend against adaptive AI threats. This points to a direction of building robust and reliable systems considering emerging risks.

\item 
Green terms center on specific attack methodologies (deep learning, adversarial examples, backdoor attacks, feature extraction, explainable AI) that leverage loopholes in AI models. Researchers examine vulnerabilities and propose technical solutions to enhance model transparency and robustness.

\item Yellow keywords highlight computational techniques (attack detection, cyber security) being developed for timely identification of SE activity at scale. 

\end{itemize}
By analyzing trends across these focus areas, this research aims to provide a holistic overview of the evolving SE threat landscape in relation to progressing digital technologies. It also seeks to identify open challenges and outline opportunities for developing responsible, multi-pronged defensive solutions through rigorous interdisciplinary collaborations. This work helps guide secure, trustworthy integration of advancing computational capabilities with social systems.
\section{Research progress}\label{sec3}

\begin{figure}[htbp]
    \centering
    \includegraphics[width=0.9\linewidth]{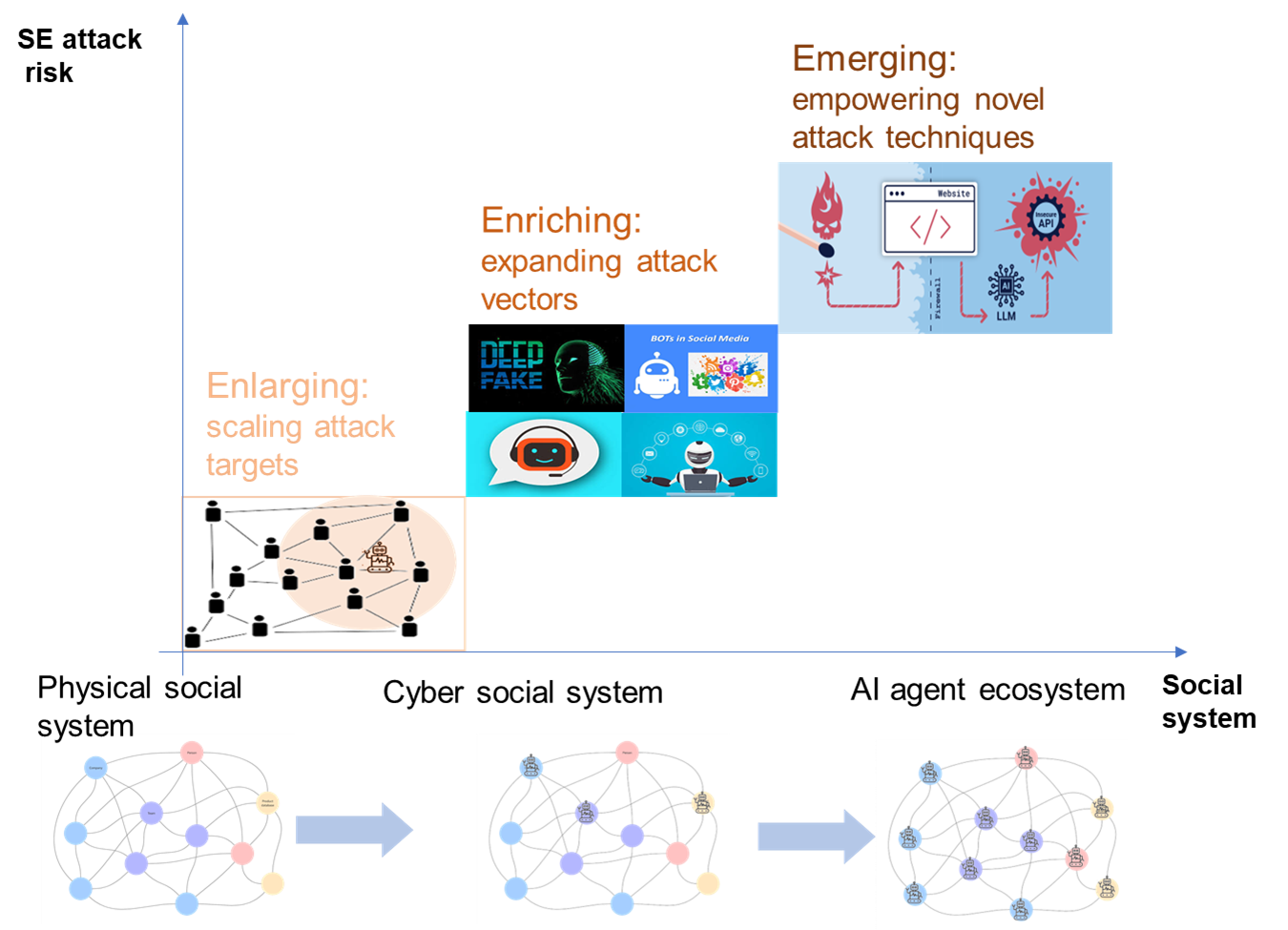}
    \caption{The 3E evolving landscapes of social engineering parallel to AI techniques across three phases of social system.}
    \label{3E}
\end{figure}

This section characterizes the history and current state of social engineering attack methods and scenarios.
The evolution of social engineering attacks has seen significant 3E trends(enlarging, enriching and emerging) parallel to advances in AI techniques across three phases of social system, as shown in Fig.~\ref{3E}. Early SE approaches primarily relied on interpersonal interaction, leveraging human psychology and lacking technological augmentation. With emerging data-driven methodologies and artificial intelligence, SE tactics have grown increasingly sophisticated. As cyber connectivity emerged via online social platforms, SE attacks manifested an Enlarging trend, scaling traditional human-driven techniques to reach broader targets through automation. Subsequently, SE attacks began an Enriching process by enabling personalized, modularized campaigns exploiting networked user dynamics. At the latest stage of sophisticated autonomous agent technologies, their involvement in manipulation shows signs of Emerging novel forms of assault. 

To contextualize the technical progression, we analyze SE techniques through two developmental stages: (1) human-based traditional SE, (2) AI-empowered SE. 
The literature are analyzed according to major methodological shifts in SE attacks that are commensurate with enabling technological advances, particularly those related to AI. 
By benchmarking methodological shifts, our analysis contextualizes the rising sophistication of SE commensurate with enabling technologies. The outlined historical trajectory and analytical framework serve to both retrospectively organize existing works and prospectively anticipate future developments.

\subsection{Evolving landscapes of social engineering: from enlargement to emergence }

A common taxonomy in literature is to divide social engineering attacks into human-based and computer-based~\cite{damle2002social,mohd2012generic,maan2012social}. 
In human-based attacks, the attacker executes the attack in person by interacting with the target to gather desired information. Thus, they can only influence a limited number of victims.

\subsubsection{Human-based attacks: Traditional social engineering techniques}

This section characterizes traditional social engineering techniques reliant on human interaction and manipulation to deceive targets and obtain sensitive information. Human-based attacks are performed through relationships with the victims to play on their psychology and emotion~\cite{salahdine2019social}.These attacks are the most dangerous and successful attacks as they involve human interactions. Examples of these attacks are pretexting, phishing, vishing, impersonation, and others:
\begin{enumerate}
\item 
\textbf{Pretexting} 
Pretexting fabricates scenarios pressuring disclosure under false pretenses, commonly targeting client data from industries including finance and utilities. ~\cite{dance2007pretexting}. During pretexting, the attack actor will often impersonate a client or a high-level employee of the targeted organization.
\item
\textbf{Phishing}  
Attackers send spoofed emails appearing to be from legitimate organizations or individuals, with malicious links or soliciting sensitive information. Phishing emails often impersonate banks, e-payment services, social media platforms requesting credentials like usernames, passwords, credit card numbers by clicking on included links~\cite{desolda2021human}.
\item
\textbf{Vishing} 
Attackers pretend to be representatives of legitimate organizations over phone calls, attempting to fraudulently obtain personal information or sensitive data from targets~\cite{griffin_vishing_2008}. They may claim to be from banks, government agencies, IT support teams etc. and leverage social engineering to trick targets into disclosing passwords, account numbers, social security numbers for fraudulent purposes.
\item
\textbf{Impersonation} 
Impersonation assumes false identities. Traditionally, impersonators take on identities of management, staff or clients. The goal is to infiltrate networks or acquire unauthorized access to data by establishing rapport~\cite{salahdine2019social}. 
\item
\textbf{Others}
Other techniques include \textit{baiting}, which tempts disclosure through enticing offers. \textit{Tailgating} physically infiltrates access-controlled areas by trailing authorized personnel unnoticed. \textit{Quid pro quo} proposes service-for-data exchanges, commonly technical support impersonating IT staff.

\end{enumerate}

The foundations of traditional social engineering techniques were examined in early literature. 
Twitchell (2009)~\cite{twitchell_social_2009} outlined typical social engineering tactics in the pre-computing era like confidence schemes, impersonation and examined applicable countermeasures.
Hadnagy (2010)~\cite{hadnagy_social_2010}introduced the art of social engineering and common techniques. 
Lohani (2019)~\cite{lohani_social_2019}introduced the concept of social engineering, its goals, methods like pretexting and psychological manipulation, establishing "hacking the human" as the core vector.
This section introduces foundational techniques established prior to computer-based SE attacks, laying indispensable groundwork for subsequent exploration. 

\subsubsection{Transition to computer-based attacks: Automation and scalability}

As cyber connectivity emerged via online social platforms, SE attacks manifested an "Enlarging" trend, scaling traditional human-driven techniques to reach broader targets through automation.

\paragraph{Enlarging: scaling attack targets through online medium}

With the rise of Internet and social networks, social engineering attacks began leveraging online platforms to widen their reach. Perpetrators induced victims to disclose personal or financial data via fabricated identities and phishing emails. Attack medium diversified to include but not limited to emails, instant messages and social media platforms.
With the rapid development of AI, social engineering attacks demonstrated more precise and covert characteristics. Leveraging big data analytics, attackers designed more effective inducement strategies by better understanding victims' behavioral patterns.

Several studies have explored the threats of SE attacks ,especially the cybersecurity. Murphy et al.(2007)~\cite{ murphy_phishing_2007} outlined threats posed by phishing, pharming and vishing to cybersecurity.
Park and Seo (2007) ~\cite{park_study_2007}studied certification approaches to preventing information leakage in phishing, vishing and smishing attacks.
Alazri (2015)~\cite{alazri_awareness_2015}  discussed social engineering paradigms and challenges in the era of information revolution. Mattera et al.(2021)~\cite{mattera_social_2021}predicted social engineering would become a major information security threat. 

As digital networks and vast data repositories became ubiquitous, new opportunities emerged for social engineers to refine traditional techniques through data-driven customization and scale.
\begin{enumerate}
\item
\textbf{Pretexting} 
With access to comprehensive personal profiles and relationships online, pretexting attackers can craft highly tailored fabricated pretenses likely to elicit information without raising suspicion. Tricking victims becomes easier as identities and interests proliferate in digital dossiers. However, such empowerment also raises complex ethical dilemmas around privacy and consent.

According to Zotina (2022)~\cite{ zotina_pretexting_2022}, criminals employ varied pretexting methods in criminal investigations. However, pretexting poses challenges regarding privacy, ethics, and consent~\cite{bennett2009ethics,dance2007pretexting}. Pretexting circumvents consent requirements by eliciting information through fabricated pretenses, raising complex questions around privacy, autonomy, and information control. 
\item
\textbf{Phishing}  

Entering the digital era, phishing benefited immensely from massive transaction and social media records revealing contextual insights into targets. Malicious actors could craft hyper-personalized phishing lures that evaded suspicion by mimicking familiar organizations and leveraging open source data to establish rapport. 
Sheng et al.(2009)~\cite{sheng2009empirical} studied the effectiveness of phishing blacklists in detecting phishing websites. They analyzed features of websites in blacklists and compared them with benign websites to identify differences that blacklisting techniques rely on.
Desolda et al.(2021)~\cite{desolda2021human} conducted a systematic literature review to analyze human factors in phishing attacks. They identified technical and socio-technical factors that influence individual susceptibility to phishing, including education, personality traits and deception techniques.
\item
\textbf{Vishing} 

As social networking increasingly transpires in digital medium, more personal information is being collected and misused. The growing availability of individual data online has facilitated socially engineered attacks leveraging people's information.

Research on social engineering techniques, particularly vishing, has continued to deepen. Jones et al. (2021)~\cite{jones_how_2021} studied the persuasion principles used by social engineers in carrying out vishing attacks, finding that selectively using vague or sensitive information can more effectively persuade victims. Maseno (2017)~\cite{maseno_vishing_2017} proposed a model for detecting mobile user vishing attacks.
Armstrong et al. (2023) ~\cite{armstrong_how_2023} compared perceptions of caller honesty in vishing attacks using either sensitive or innocuous requests. Ashfaq et al. (2024) ~\cite{roy_defending_2024} reviewed vishing attacks and summarized prevention and mitigation techniques.
\item
\textbf{Impersonation} 

Burgeoning digital dossiers detailing identities, relationships and activities have empowered impersonators to simulate authentic personas more convincingly. 
Tu et al. (2021)~\cite{tu2021reinforcement} proposed a reinforcement learning model to detect impersonation attacks in device-to-device communications for future networks.
Sheng et al.(2018)~\cite{bala2018impersonation} analyzed security of a certificateless encryption scheme and found opportunities for impersonation attacks by adversaries reconstructing public keys from leaked system parameters.
\end{enumerate}

\paragraph{Enriching: expanding attack vectors through deep learning techniques}

With the rise of deep learning, assaults began an enriching diversification through increasingly specialized model applications. 
Technological developments have enabled the automation and large-scale deployment of social engineering attacks.

\textbf{Personalized attacks via user profiling} Latent feature models powered by large-scale user data allow profiling interests, preferences and behaviors for hyper-targeted phishing.

\textbf{Synthetic media generation using Generative adversarial networks} GANs are a powerful generative model consisting of a generator and discriminator that adversarially train together to enable the generator to produce convincing fake samples. Generative adversarial networks synthesize convincing fake multimedia, including deepfakes that realistically impersonate identities. 

\textbf{Deceptive content via natural language processing models(NLP)} NLP techniques have been widely applied to generate deceptive contents, including fraudulent text generation and semantic manipulation. Neural language models produce deceptive or manipulated text at scale. Automated writing masks artifices under layers of human-like language.

\textbf{Optimized interactions through reinforcement learning} Reinforcement learning trains chatbots via feedback to continually refine social deceptions over multiple episodes. Dialog systems maximize yields by learning optimal engagement strategies.

Collectively, advances in AI have supercharged social engineering’s once human-constrained abilities and enabled new social engineering attack methods using emerging vectors like deepfakes, virtual assistants, social media bots and chatbots.


\textbf{Deepfake}

Deepfake technology leverages generative adversarial networks and other deep learning models to replace the features in an image or video with those of another image, generating highly realistic fake content. The advancement of deepfake technology has made the creation of high-quality fake media simple and inexpensive, further exacerbating its application in social engineering attacks.

Recent advances in machine learning have enabled the emergence of deepfakes, hyper-realistic digital videos manipulated using artificial intelligence to depict individuals engaging in fabricated acts (Westerlund, (2019)~\cite{westerlund_emergence_2019}; Albahar and Almalki, (2019)~\cite{
albahar_deepfakes_2019}). Deepfakes pose societal risks by impairing the authenticity of visual evidence and disseminating manipulated content (Pantserev, (2020)~\cite{jahankhani_malicious_2020}; Katarya and Lal, (2020)~\cite{katarya_study_2020}). Deepfakes threaten news reliability, national security, and individual privacy by enabling propaganda and impersonation. 

Studies~\cite{jahankhani_malicious_2020, ranganathan_review_2022,katarya_study_2020}analyzed the threats of generating hyper-realistic audio-visual deepfake content for the purpose of manipulating public opinion and undermining political stability.
Additionally, existing studies ~\cite{jones_artificial_2020,botha_fake_2020} suggested deepfake technology elevates information security risks by enabling the fabrication of news and personal details, affecting public judgment formation.
Karasavva et al.(2021) ~\cite{karasavva_real_2021} discussed the impacts of deepfake pornography on privacy and gender equality, transforming it into a new means of online harassment.
Langa et al.(2021)~\cite{langa_deepfakes_2021} explored legislative challenges as deepfake content is difficult to verify, bringing new regulatory issues with emerging attack vectors.
Galyashina et al.(2022)~\cite{galyashina_protection_2022} reported the usage of deepfake in megascience projects to potentially deliver malware or exfiltrate confidential data via social manipulation.

Overall, deepfake empowers SE attacks in three major ways.
Firstly, it enables the generation of fabricated information in more persuasive audio-visual forms to interfere public opinion steering.
Secondly,it increases content stealthiness and expanding the scope of unsuspecting users under attack, exacerbating online harassment issues;
Thirdly, it challenge detection and governance with traditional methods due to the deceptive nature, posing novel threats for defensive work.
Deepfake was identified as an emerging frontier in SE attacks, signifying the importance of researching its societal risks from technological and policy perspectives.

\textbf{Virtual assistant}

Virtual assistants (VAs), enabled by advances in natural language processing and speech recognition technologies, have gained widespread adoption in recent years for assisting users with daily tasks via conversational interfaces~\cite{chung_intelligent_2018}. While offering frictionless interaction, VAs also present emerging security and privacy concerns due to their architecture which relies on cloud computing platforms to power functionality~\cite{maedche_ai-based_2019}.

Several studies have demonstrated successful malicious attacks targeting VAs~\cite{maedche_ai-based_2019,
mitev_alexa_2019}, indicating their increasing appeal to threat actors. 
Zhang et al.(2019)~\cite{zhang_dangerous_2019} analyzed security risks of third-party skills on VA platforms and proposed mitigation approaches. Risks included malicious skills bypassing review processes.
An et al.(2018)~\cite{an_malware_2018} applied malware detection techniques to identify anomalous behavior in VAs indicative of potential compromise.
More worryingly, attacks are evolving to employ remote vectors and more sophisticated techniques. At the same time, personal data extraction from unsecured VAs has been shown feasible~\cite{bolton_security_2021}. Evaluations of key data protection legislation also found limited protections for VA users .

Researchers have also summarized the key security and privacy challenges facing VAs, such as concerns around data protection and unwanted triggering~\cite{bolton_security_2021}.
Regarding system security, vulnerabilities have been exposed in third-party application review processes enabling malicious skill approval~\cite{zhang_dangerous_2019}. 
Dreyling et al. (2021)~\cite{dreyling_cyber_2021} conducted a cyber risk analysis of a government VA digital service using the FAIR model to identify vulnerabilities.
Kumar et al.(2021)~\cite{kumar_ai_2021}introduced a computational trust model based on AI to evaluate a VA's trustworthiness.
Applications of VAs for security were proposed. Pandit et al.(2022)~\cite{ pandit_fighting_2020, pandit_combating_2023} proposed leveraging a virtual assistant to mediate user interaction with communication systems like phones to combat robocalls and voice spam more effectively.

In summary, as VAs gather increasing user data within immature regulatory landscapes, security and privacy remain open challenges deserving further interdisciplinary exploration. Robust defenses and well-informed user comprehension will be critical to realising VAs' societal potential while mitigating associated risks.

\textbf{Social media bots}

Social media platforms host large numbers of automated accounts, or "bots", which utilize computer scripts to emulate human behaviors and manipulate discussions.
Social media bots have increasingly garnered research attention due to their potential to distort online conversations and public opinions at scale. Several studies have employed literature reviews and empirical analyses to characterize bot activities.
Notable work by Ferrara(2015)~\cite{ferrara_manipulation_2015} summarized the technical and societal challenges posed by bot misuse, such as spreading misinformation, opinion manipulation, and coordinated inauthentic campaigns.
Social engineering attacks enabled by social media bots can broadly be categorized into three classes based on prior empirical work: information manipulation, emotional influence, and opinion control.

Information manipulation attacks aiming to sway public opinions through disseminating falsehoods have been documented. Ferrara et al.(2020)~\cite{ferrara_characterizing_2020} analyzed coordinated bot campaigns on Twitter during the 2020 US elections, finding evidence of coordination albeit on a smaller scale than humans. Chang et al. ~\cite{chang_social_2021}reviewed the role of bots, misinformation, and platform interventions during COVID-19 through 2020.

Computational modeling has demonstrated emotional influence attacks that leverage bots to amplify user emotions.  Wagner et al. (2012)~\cite{wagner2012social} showed through agent-based simulation that bots targeting emotional content can increase anger online. Their experiments showed that bots targeting emotional content can amplify anger in online discussions. Weng and Lin (2022) ~\cite{weng_public_2022} conducted social network analyses of Twitter data coinciding with peaks in the Wuhan lab leak conspiracy, finding bot-driven emotional manipulation. 

At large scales, opinion control attacks facilitating social fraud have been exhibited, like Paquet-Clouston et al.(2017) ~\cite{paquet-clouston_can_2017} revelation of IoT botnet manipulation on Twitter. They demonstrated how an IoT botnet conducted social network manipulation on Twitter through social media fraud services selling artificial likes, followers and views. 
Inventories of organized campaigns were provided~\cite{bradshaw_troops_2017}. Deb et al.(2019)~\cite{deb_perils_2019} conducted an analysis of social media data from the 2018 US midterm elections to detect potential election manipulation.

Phishing attacks guided by bot taxonomies distinguishing commercial and surveillance bots were defined~\cite{gorwa_unpacking_2020}.
Gorwa and Guilbeault~\cite{ gorwa_unpacking_2020}  provided a typology of bots to guide related research and policy-making. They identified five main bot types based on function and intent, including commercial bots, political bots, surveillance bots, spam bots, and deepfakes/synthetic media bots.

Studies have also advanced detection and defenses, such as Orabi et al.(2020) ~\cite{orabi2020detection} summarization of detection techniques and Benigni et al.(2019)  ~\cite{agarwal_bot-ivistm_2019} coordinated behavior identification framework. Orabi et al.(2020)~\cite{orabi2020detection}performed a systematic literature review of bot detection techniques on social media. They analyzed 55 papers and identified three main categories of approaches: content-based, graph-based, and hybrid methods. Content-based methods analyze profile attributes, linguistic features, and temporal patterns in posts. Graph-based techniques leverage social connections and interaction networks. Hybrid models integrate multiple information sources.
Benigni et al.(2019)~\cite{ agarwal_bot-ivistm_2019} proposed a framework called Bot-ivistm to assess information manipulation by social bots through network and community structure analytics. They demonstrated its application to understand coordinated campaigns on Twitter and Gab.
Bradshaw and Howard (2017)~\cite{ bradshaw_troops_2017} provided a global inventory of organized social media manipulation campaigns identified through open-source information and technical data. This work provides a basis for understanding prevalent SE attack patterns leveraging social bots through empirical analyses.

In summary, the existing studies have advanced the understanding of bot behaviors through systematic analyses and empirical case studies. Detection and mitigation of bots and bot-enabled online harms remain a challenging open problem with many socio-technical considerations. Future research directions include tracking evolving bot strategies, modeling information flows in hybrid human-bot networks, and developing responsible governance frameworks.

\textbf{Chatbot}

Chatbots can be manipulated to provide false information, leading users to click on links containing malware or fraudulent websites. Additionally, insecure AI chatbots may be used to spread misinformation, manipulate public discourse, and even steal users' personal privacy. Existing study has ~\cite{ye2020chatbot}  reviewed chatbot security and privacy challenges arising from their widespread use in personal assistant scenarios. Technical issues like data leaks, functional attacks and lack of accountability were covered.
Chatbot-enabled SE attacks can broadly be categorized into four types based on the literature.

Information manipulation attacks aim to influence public opinions and viewpoints through misleading or spreading misinformation via chatbots, as explored in experiments testing political influence and analyses of potential manipulation risks. Huang-Isherwood et al.(2024)~\cite{ huang-isherwood_human_2024} conducted an experiment testing how chatbot manipulations can influence people's political self-concepts. Their results provided guidance on bot manipulation detection heuristics. Carroll et al.(2023)~\cite{ carroll_characterizing_2023} characterized potential manipulation risks from AI conversational systems and mitigation approaches. Rayhan~\cite{ rayhan_dark_nodate} discussed chatbot manipulation techniques from an adversarial perspective.

Behavioral impact attacks attempt to affect user psychology and conduct following chatbot interactions through examining self-disclosure effects on mental models and blame attribution. 
Ho et al.(2018)~\cite{ ho_psychological_2018} investigated the psychological, relational and emotional effects of self-disclosure after conversations with a chatbot. Crolic et al.(2022)~\cite{crolic_blame_2022}explored how anthropomorphism and anger in customer-chatbot interactions can influence blame attribution through empirical studies. De Gennaro et al.(2020)~\cite{de_gennaro_effectiveness_2020} evaluated an empathic chatbot's effectiveness in combating adverse social exclusion effects on mood.

Phishing attacks utilize chatbots impersonating enterprises or individuals for online scams, with proposed defenses against social networking phishing. Yoo and Cho (2022) ~\cite{yoo2022icsa} proposed an intelligent chatbot security assistant (ICSA) using text CNN and multi-phase defenses to protect against SNS phishing attacks.

Service impact attacks seek to diminish service quality or disrupt normal operations of businesses through chatbot interactions, addressing negative consumer perceptions. Roy and Naidoo  (2021)~\cite{roy_enhancing_2021} assessed the impact of chatbot presentation styles and time orientation on effectiveness in business contexts. Mozafari et al. (2021)~\cite{mozafari_resolving_2021} leveraged selective chatbot self-disclosure to mitigate negative user perceptions in different service scenarios.
Shumanov and Johnson (2021) ~\cite{shumanov_making_2021} explored personalizing chatbot conversations for enhanced experience.

In summary, existing studies have empirically analyzed key technical, behavioral and application factors regarding chatbot security, effectiveness and potential manipulation risks, providing guidance for responsible chatbot design and deployment. Future work can consider evolving chatbot adversarial behaviors and large-scale impact assessments.

\paragraph{Emerging:
empowering novel attack techniques through large language models
}

Recently, involvement of LLM portends the "Emerging" of qualitatively novel assault forms. Systems demonstrating generative language proficiency intrinsically risks.
By innovating in model architectures, training optimization, learning paradigms and other aspects, LLM built upon the foundation of neural networks and leveraged computational resources and big data advantages to break through the limitations of traditional neural networks in terms of scale, learning capability and broad applicability\cite{yu_data_2024a, yu_navigating_2024}. The key technological breakthroughs include:
\begin{itemize}
\item

In model architectures, transformer self-attention and other mechanisms were used to design larger-scale models with parameter sizes usually in the millions or even over billions, such as GPT-3's 1.75 billion parameters.

Such large models enable generating hyper-realistic deceptive contents at scale.
\item
In training optimization, the pre-training and fine-tuning transfer learning paradigm was realized to promote cross-task learning. GANs and other generative models were trained with reinforcement learning to gain the ability of generating new content, including fake identities, dialogs or documents for social manipulation. 
\item
In learning paradigms, large-scale self-supervised learning was conducted to excavate deep representations in corpora. Prompting engineering was introduced to improve interactive learning flexibility and efficiency, facilitating the generation of deceptive but human-like responses for social engineering attacks.
\end{itemize}

LLM systems may enable new forms of social engineering attacks if vulnerabilities are introduced during the development lifecycle. Key stages include data preparation, training, deployment, application, and updating. 
At each stage, vulnerabilities could be intentionally or unintentionally created that amplify social harms, as outlined in Tab.\ref{tab:LLM_SE}.

\begin{table}[htbp]
\centering
\caption{LLM enabled SE attack techniques}
\label{tab:LLM_SE}
\begin{tabular}{p{1.5cm}p{2cm}p{2.5cm}p{1.5cm}}

\toprule
Sub-stages  & How to Empower Social Attacks & Enhancing Techniques  & Reference  
\\ 
\hline
\multirow{2}{1.5cm}{Data Collection and Preparation} & \multirow{2}{2cm}{Empower building trust and   manipulating information} & Data Poisoning Attack  & ~\cite{panda_teach_2024,zhang_human-imperceptible_2024,yang_watch_2024,he_data_2024} 
\\ 
\cline{3-4} 
 &  & Biases in Data Preparation and Annotations  & ~\cite{navigli_biases_2023,barberio_large_2022,ray_chatgpt_2023,kolisko_exploring_2023,mishra_llm-guided_2024,taubenfeld_systematic_2024,chen_humans_2024} 
 \\ 
 \hline
\multirow{2}{1.5cm}{Model Training}  & \multirow{2}{2cm}{Automating large-scale attacks} & Model Poisoning Attack & ~\cite{zhang_human-imperceptible_2024,zou_poisonedrag_2024}
\\ 
& & Data Reconstruction  Attack  & ~\cite{elmahdy_deconstructing_2023,wang_pandoras_2024,li_drattack_2024} 
\\ 
\hline
\multirow{4}{1.5cm}{Model Deployment} & \multirow{2}{2cm}{Automating large-scale attacks} & Adversarial Examples  & ~\cite{raina_is_2024,zou_universal_2023,yao_llm_2023,li_fmm-attack_2024} 
\\ \addlinespace
& & Jailbreak & ~\cite{chu_comprehensive_2024,xu_llm_2024,chang_play_2024,huang_catastrophic_2023,wei_jailbroken_2024}  
\\ 
& \multirow{2}{1.5cm}{Extracting
privacy data}   & Data Theft from AI Memory  & ~\cite{greshake_not_2023,panda_teach_2024,sai_generative_2024}  
\\ 
&  & Membership Inference Attack &~\cite{fu_practical_2023,kandpal_user_2023,duan_membership_2024} 
\\ \addlinespace
\hline
\multirow{4}{1.5cm}{Model Application}  & \multirow{2}{2cm}{Information manipulation} & \multirow{2}{2cm}{Trustworthiness of AIGC} & ~\cite{wang_decodingtrust_2024,lu_gpt-4_2024,haufglockner_self-supervised_2023,greevink_ai-powered_2023}  
\\ \addlinespace
 &   &   &  
 \\ 
& Extracting privacy data & Privacy Leakage in Interacting with AI  & ~\cite{agarwal_investigating_2024,kumar_ethics_2024} 
\\ 
& Automating large-scale attacks  & Prompt Injection  & ~\cite{liu_automatic_2024,pasquini_neural_2024,zhan_injecagent_2024,bagdasaryan_abusing_2023,liu_prompt_2023}  
\\ 
\hline
\multirow{5}{1.5cm}{Model updating}  & \multirow{5}{2cm}{Automating large-scale attacks}  & Code injection attack  & ~\cite{aryal_intra-section_2024,noman_code_2023}  
\\ 
 &  & Parameter stealing attack   & ~\cite{oliynyk_i_2023,carlini_stealing_2024} 
 \\ 
&  & Edge node intrusion  & ~\cite{yang_efficient_2023}  
\\ 
 & & Update mechanism cracking  & ~\cite{shourya_comparative_2023}  
 \\ 
 &  & Backdoor implantation  & ~\cite{huang_composite_2024,he_transferring_2024}  
 \\ 
\bottomrule
\end{tabular}
\end{table}

\textbf{Data Collection and Preparation}

During data collection and preparation, there exist attack methods through which training regimes could potentially enable wide-scale social engineering by LLMs. Firstly, data poisoning attacks, where adversarially manipulated examples introduce biases by exposing models to simulated techniques like phishing messages without context or fact-checking. Secondly, annotation biases, as human annotators influenced by preconceptions may misclassify "successful" manipulation strategies observed online, normalizing insincere persuasive styles.

Data poisoning techniques have been explored as potential attacks on language models. Poisoned training data could induce biases or toxicity within models~\cite{zhang_human-imperceptible_2024}. 
Biased samples could steer models toward harming groups. Existing studies profiled ChatGPT limitations ~\cite{ray_chatgpt_2023}and challenges of preprocessing with generative systems ~\cite{barberio_large_2022}. Zhang et al.(2024)~\cite{zhang_human-imperceptible_2024} analyzed how retrieval poisoning allows compromising LLMs via legitimate-appearing but manipulated outputs from powered applications.He et al.(2024)~\cite{he_data_2024} explored data poisoning attacks against in-context learning. 
Yang et al.(2024)~\cite{yang_watch_2024} demonstrated backdoor injection enabling reasoning manipulation or output controls.

Early research explored bias origins~\cite{barberio_large_2022} and how attacks inject backdoors~\cite{yang_watch_2024}.
Navigli et al.(2023)~\cite{navigli_biases_2023} explored the root causes of LLM biases and cataloged a range of specific bias types, such as gender, racial, and ideological biases. 
Ray et al.(2023)~\cite{ray_chatgpt_2023} provided a holistic introduction to ChatGPT, including its key challenges of biases and ethical concerns and future prospects. 


\textbf{Model Training}

During model training stage of LLMs, there exist two primary avenues through which adversarial attacks may facilitate social engineering at scale via LLMs. Firstly, model poisoning attacks wherein adversaries strategically introduce corrupted training examples to subtly bias model representations. For instance, poisoning aimed at associating typically neutral topics with fear, uncertainty or deception, potentially enabling generated content to similarly manipulate without detection.
Secondly, data reconstruction attacks whereby identities within anonymized datasets used for LLM training become identifiable, exposing privacy-sensitive attributes. Such techniques could theoretically power population-scale impersonation or customized misinformation diffusion through personalized synthesized profiles.


Data reconstruction attacks leverage the knowledge unintentionally learned by the models to reconstruct parts or the whole training data, leading to privacy leakage~\cite{elmahdy_deconstructing_2023,yu_swdpm_2023}. By exploiting the hidden representations encoded in the models, reconstruction attacks can extract private information from the training samples.
Elmahdy et al.(2023) ~\cite{elmahdy_deconstructing_2023} proposed a novel "Mix And Match" attack that can reconstruct the training data used for text classification models and highlighted the privacy risks associated with data reconstruction attacks.
Wang et al.(2024)~\cite{wang_pandoras_2024} investigated the problem of training data leakage in open-source large language models. 
Li et al.(2024)~\cite{li_drattack_2024} introduced a powerful attack called "DrAttack" that can bypass the safety mechanisms of large language models by decomposing and reconstructing prompts, leading to unintended and potentially harmful outputs.

For model poisoning attacks, adversaries subtly manipulate the models' internal parameters through gradient information or other insights gained from the training process~\cite{zhang_human-imperceptible_2024,zou_poisonedrag_2024}. This is typically achieved by inserting backdoors or biases during model updating.
Zou et al.(2024)~\cite{zou_poisonedrag_2024} explored a new type of attack "knowledge poisoning" that targets retrieval-augmented generation in large language models, leading to harmful outputs during text generation.


\textbf{Model Deployment}

At the model deployment stage, several attack techniques have been shown to enable novel forms of social engineering.
Specifically, adversarial examples pose risks, as judicious inputs could induce models to generate misleading outputs propagating disinformation virally.
Jailbreaking attacks attempt to circumvent access controls and coerce deployed models into disclosing sensitive knowledge or manipulating unrelated tasks.
Data extraction from model memory could aid personalized deepfakes or precision social engineering by reconstructing attributes of individuals in proprietary training datasets.
Membership inference attacks aim to determine dataset representation of specific individuals, potentially enabling targeted social attacks.

Adversarial examples research demonstrates how perturbations can mislead models into generating deceptive or harmful outputs that could influence human opinions or decisions.
Raina et al.(2024)~\cite{raina_is_2024} investigated the robustness of using LLMs as judges for zero-shot assessment tasks.
Zou et al.(2023)~\cite{zou_universal_2023} proposed universal and transferable adversarial attacks that can be applied to aligned language models.
Yao et al.(2023)~\cite{yao_llm_2023} demonstrated that random token prompts can elicit hallucinations from LLMs, suggesting hallucinations may be a form of adversarial examples.
Li et al.(2024)~\cite{li_fmm-attack_2024} introduced a flow-based multi-modal adversarial attack targeting video-based LLMs.

Other works have examined "jailbreaking" attacks, whereby an adversary manipulates an LLM into generating undesirable content without authorization.
Chu et al.(2024)~\cite{chu_comprehensive_2024} provided a comprehensive assessment of jailbreak attacks against LLMs.
Xu et al.(2024)~\cite{xu_llm_2024} conducted a comprehensive study on LLM jailbreak attacks and defense techniques.
Chang et al.(2024)~\cite{chang_play_2024} introduced an indirect jailbreak attack on LLMs using a guessing game with implicit clues.
Huang et al.(2023)~\cite{huang_catastrophic_2023} explored a catastrophic jailbreak attack on open-source LLMs by exploiting the generation process.
Despite efforts to align LLMs with human values, widely used LLMs such as GPT, Llama, Claude, and PaLM are susceptible to jailbreaking attacks, wherein an adversary fools a targeted LLM into generating objectionable content.

Additional research has profiled privacy risks, such as data theft~\cite{panda_teach_2024,sai_generative_2024} or enabling membership inference attacks.
For example, Greshake et al. (2023)~\cite{greshake_not_2023}, by strategically injecting prompts into data likely to be retrieved during inference, adversaries can exploit applications to perform arbitrary code execution, data theft, and manipulate functionality. 
Membership inference is able to reveal privacy dataset membership information, and thereby judge personal preferences, behaviors, etc~\cite{kandpal_user_2023}. Fu et al. (2023)~\cite{fu_practical_2023} proposed a new membership inference attack, showing privacy leakage remains a challenge for LLMs trained on sensitive data. Duan et al. (2024)~\cite{duan_membership_2024} evaluated whether traditional membership inference attacks work on pre-training data of large language models.

\textbf{Model Application}

During the application phase of LLM, several emerging attack methods pose risks for enabling advanced SE attacks.
LLMs' proficiency in generating synthetic yet authentic-sounding textual outputs threatens to undermine public discourse. By propagating machine-generated misinformation invisible to most users, adversaries could covertly manipulate opinions and narratives at scale. Secondly, privacy leakage vulnerabilities in conversational AI systems endanger individuals by potentially revealing sensitive attributes via response profiling. Finally, the malleability of generative models' outputs through priming language cues enables subtle yet powerful result manipulation.

Several studies have explored both opportunities and risks associated with deploying advanced generative models like LLMs.
To undermine trust in AI-generated texts, adversaries subtly implant inaccuracies or misleading claims within model outputs to manipulate public discourses and steer user perceptions~\cite{wang_decodingtrust_2024}. Several studies have explored the emerging issues surrounding AI-generated content (AIGC). Wang et al.(2023)~\cite{ wang_survey_2023} provided an overview of AIGC, including its development history and applications. Challenges in ensuring information quality and mitigating biases were also discussed. Basyoni and Qadir~\cite{basyoni_ai_2023} investigated AIGC's implications for public safety through a case study. Grimme et al.(2023)~\cite{grimme2023lost-fakenews} also studied how LLMs may be utilized to generate synthetic yet coordinated disinformation campaigns on social platforms, and proposed techniques to trace the digital fingerprints of such machine-driven messaging.

By eliciting private attributes from conversations, privacy leakage attacks emerge through response analysis during natural interactions. Personal details may also be extracted for unauthorized secondary purposes~\cite{agarwal_investigating_2024}. Liu et al.(2023)~\cite{ liu_risk_2023} analyzed risks of AIGC in market regulation, such as information disclosure and attribute inference. Chen and Tian~\cite{chen_challenges_2023} took a Marxist political economy perspective to review AIGC's challenges, including data legitimacy, biases, incentive mechanisms and learning behavior oversight. Xuan ~\cite{guo_aigc_2023, xuan_risks_2023} summarized AIGC risks in the judicial system, involving information authenticity and interpretable scrutiny of models.
Key issues included data sourcing legality, biases, unfairness from incentivization, labeling errors, and impacts on artificial general and superintelligence. 
Collectively, these studies explored emerging problems from AIGC application, focusing on domains like public safety, market rules and jurisprudence.

Prompt injection techniques pose risks by enabling the manipulation of model outputs and downstream audiences. Toxic, harmful or propagandistic prompts have been shown to guide text generation in LLMs~\cite{zhan_injecagent_2024,bagdasaryan_abusing_2023}. Several works have sought to systematically characterize these attacks.
Liu et al.(2023)~\cite{liu_prompt_2023} proposed a framework for analyzing prompt injections and defenses.
Liu et al.(2024)~\cite{liu_automatic_2024} introduced an automated gradient-based method for generating highly effective universal injections requiring just five training samples, outperforming baselines.
Pasquini et al.(2024)~\cite{pasquini_neural_2024} defined "Neural Execs" - autonomously generated execution triggers bypassing protections via flexibility.
In general, prompt attacks work by converting adversarial textual content into an injection prompt that induces LLMs to output adversarial samples matching an attack goal (Liu et al., 2023). Effective prompts typically comprise the original input or label, a task illustrating a semantically preserving perturbation, and attack guidance modifying the input at character, word and sentence levels.

Additionally, several malicious applications like WormGPT and FraudGPT for automating and scaling malicious activities also show the degree of weaponization of LLM technology.
FraudGPT gained rapid adoption by simplifying complex hacking techniques into an automated service that even non-experts could easily employ for activities such as writing malicious code, creating undetectable malware, and crafting convincing phishing emails. Compared to manual research, LLM greatly reduces the time needed to discover vulnerabilities, collect credential data from victims, learn new hacking tools, and master sophisticated cybercrimes.

At scale, automatable LLMs have reduced barriers to adversarially phishing, deceiving or advertising without authorization - highlighting the pressing need for robustness evaluations and targeted defenses against emergent sociotechnical threats. Continued rigorous study remains crucial to holistically understand and address model subversion risks.

\textbf{Model updating}

Model updating presents security vulnerabilities that may be exploited to facilitate large-scale social engineering efforts. During maintenance and revision cycles, key attack surfaces emerge. For example,
code injection attacks enabling adversaries to clandestinely alter update codepayloads, potentially implanting covert backdoors or malware. This modifies output control to propagate manipulated content and spread misinformation.
Parameter stealing attacks extracting private revision information could reveal sensitivedomain knowledge embedded within the model. Compromised personal data might be leveraged to reconstruct impersonation models for precision-targeted influence operations.
Edge node intrusions granting unauthorized access to remotely deployed instances provide vectors formanipulating outputs at scale or extracting user data with amplified sociotechnical impacts.
Circumventing authentication through update mechanism cracking risks broad dissemination offalsified narratives aimed at swaying perspectives on issues of societal import.
Backdoor implantation during revision cycles potentially introduces surreptitious biases steeringgenerative capabilities toward deception. Fine-tuned trojans may transfer across mediums topropagate influence campaigns.

For code injection attacks, adversaries may illegally modify update code to implant stealthy backdoors or malware aimed at controlling model outputs to disseminate misleading or manipulated information distorting public views. The studies studied code injection attacks in memory for malware ~\cite{aryal_intra-section_2024} and in IoT devices ~\cite{noman_code_2023}, respectively.
In parameter stealing attacks, extracting internal parameters from updates could leak privacy knowledge or build sophisticated impersonation models based on reconstructing personal details for targeted influence operations. Oliynyk et al.(2023)~\cite{oliynyk_i_2023}  provided a taxonomy for model stealing attacks and proposed categorizing different types of attacks based on goals.
Edge node intrusion enables unauthorized access to models deployed on edge devices, allowing outputs to be manipulated or data to be leaked at scale. Yang et al.(2023)~\cite{yang_efficient_2023} studied efficient intrusion detection using cloud-edge collaboration to improve performance.   

By cracking update mechanisms via vulnerabilities like auto-update flaws, attackers could broadly distribute fake news or disruptive content to sway discussions. Shourya et al.(2023)~\cite{shourya_comparative_2023} investigated dictionary attacks, where an attacker tries to determine account credentials by cycling through a list of common values. 
Huang et al.(2024)~\cite{huang_composite_2024} proposed a Composite Backdoor Attack (CBA) against LLMs. Unlike repeating trigger keys in a single component, CBA scatters triggers across components. 
He et al.(2024)~\cite{he_transferring_2024} studied transferability of backdoors across languages and confirmed that with instruction tuning, backdoors transfer successfully from the source to target languages, deceiving models.

The development of AI also risks ideological fractures, manipulation, and destabilization. Applications involving LLM, particularly conversational systems like ChatGPT and Codex, could propagate harmful societal biases at scale. While assisting legal professionals with reviewing documents, drafting legal texts, and participating in proceedings, AI integration in the legal sphere introduces novel security vulnerabilities and ethical dilemmas.
Several avenues exist for adversarially influencing ideological stances, including by selectively training on biased data, inducing membership inferences about sensitive attributes, and surreptitiously injecting malicious payloads. Defenses against such ideological threats require addressing data selection, model architectures, and what information models externally retrieve.

\subsection{Social engineering attack scenarios}

\begin{figure} [htbp]
    \centering
    \includegraphics[width=0.9\linewidth]{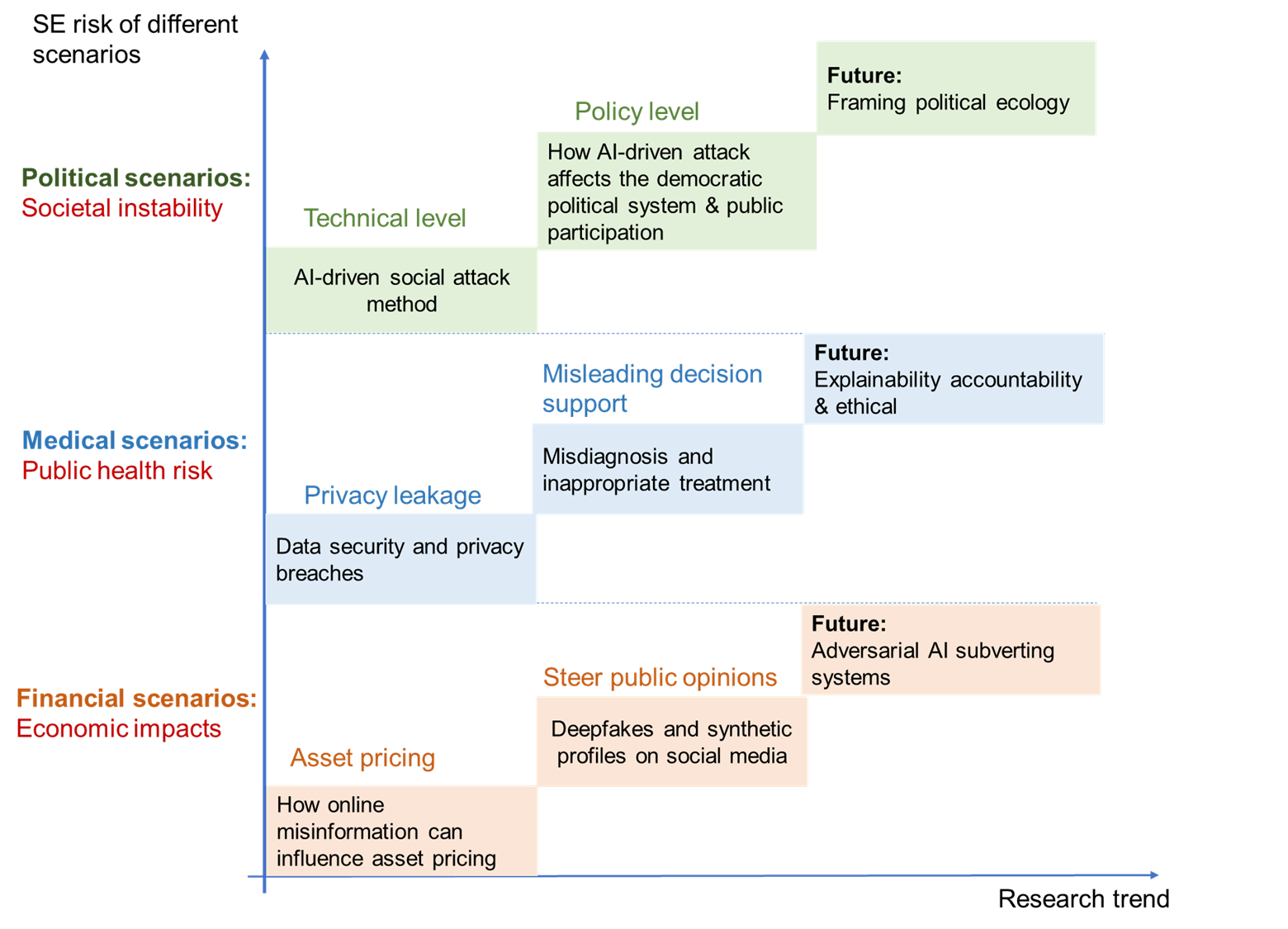}
    \caption{Research trends related to evolving social engineering attack in different scenarios.}
    \label{scenario}
\end{figure}

After analyzing SE attack tactics, Fig. \ref{scenario} outlines research trends related to evolving social engineering attack scenarios. There are three scenarios mainly discussed in this paper: financial, plitical and medical scenarios. Firstly, in economics, focus has diverged from asset pricing towards tactics surreptitiously shaping public discourse and subverting systems to indirectly steer socio-ecological conditions on a macro scale. 
Secondly, for healthcare system, concerns surrounding privacy leakage have broadened to encompass pressing issues of bias, unfairness and the need for redress mechanisms owing to personalized decision support integrated into human judgements at massive scales. 
Finally, within political applications, attention is pivoting from narrow technical considerations towards holistic policy frameworks sensitive to novel computational propaganda vectors. These research trajectories highlight the nascent, transdisciplinary nature of comprehending and preempting escalating socio-technical risks introduced by the digital transformation of society.

\subsubsection{Financial scenarios}
SE attacks in financial contexts primarily include rumor attacks, privacy leaks, and subverting control of financial systems.
Prior studies have shown that rumors can influence asset prices in capital markets~\cite{ahern_rumor_2015}. Some research found that rumors may cause temporary stock price overreactions and mispricing especially during atypical periods such as bear markets ~\cite{ he_information_2016}. Social media platforms have become major channels for rumor propagation. Studies found that artificial accounts powered by AI techniques on social networks can effectively spread rumors and manipulate public opinions similar to authentic users~\cite{he_cost-efficient_2016}, potentially impacting financial markets.

Unlike traditional rumors, AI-powered social engineering attacks are more difficult to identify and control as the generated content cannot be distinguished from reality~\cite{mishra_exploring_2023}.
This may exacerbate uncertainties among market participants and lead to more adverse consequences. In addition, the diversified media landscape poses challenges for monitoring and fact-checking rumors in a timely manner. Existing research shows media coverage itself could potentially fuel rumor transmission~\cite{yang_rumor_2014, dong_leveraging_2018}, which may similarly exist in financial context.
Importantly, rumors are prone to influencing individual behaviors and decisions, hence affecting broader domains. Some findings suggest that rumors can trigger herd panic sentiments~\cite{greenhill_rumor_2017}, negatively impacting not only financial markets but the entire economy.

On privacy leaks, attackers utilize deep learning to analyze online banking transaction records and excavate users' privacy-sensitive information for illegal use~\cite{ dong_leveraging_2018}. Attackers may also collect a large amount of user data, using GAN to generate fake user profiles for infiltration in transaction systems to steal real user data. It could even lead to large-scale blackmail scams. Attackers collect information from social media or large personal datasets~\cite{olabanji_effect_2024}(e.g. email logs, browsing histories, hard disks or phone contents), which do not necessarily constitute curse evidence, and then identify many potential targets' specific vulnerabilities customized with threatening messages.

Moreover, with the development of AI technologies, AI-driven SE attacks that subvert control of financial systems could emerge. In general, the more complex the supervised system for financial transactions, the more difficult it is to defend fully. Adversarial disturbance phenomena highlight this issue, indicating sufficiently advanced AI may inherently facilitate carefully designed attacks.



\subsubsection{Medical scenarios}
The application of AI technology in the medical field has made significant progress, including but not limited to clinical diagnosis, medical treatment, medical rehabilitation, disease prediction, and medical research. These applications not only improve the efficiency and quality of medical services but also alleviate the workload of medical staff\cite{yu_pursuing_2023,OptimizeAccessibility2023}. However, with the rapid development and widespread application of AI technology, AI-driven SE attacks have also brought a series of risks and challenges in the medical field.

The main AI-driven SE attacks in the medical field include:
misdiagnosis and inappropriate treatment caused by decision-making errors due to large model hallucinations in AI-assisted decision-making systems, data security and privacy breaches, AI model biases leading to misleading decision support and cybersecurity attacks causing decision-making failures.

If the training data is biased, the LLM system may produce misleading decision support, leading to misdiagnosis or inappropriate treatment. AI systems built on large amounts of medical data also face the risk of information leakage and misleading decisions.The study~\cite{chen_generative_2024} discusses the privacy and security challenges of generative AI in medical practice from an end-to-end perspective. It analyzes the privacy and security threats that may arise in data collection, model training, and implementation phases, and proposes corresponding risk management recommendations.

In terms of cybersecurity, the application of AI technology in the medical field may also bring security risks, including cyber attacks and data breaches. Since AI systems often rely on large amounts of patient data for learning and prediction, they may become targets of cyber attacks, threatening patient privacy and safety. 
He et al.(2020)~\cite{he_ai-based_2020} studies the AI security attack pathways for cardiac medical diagnosis systems and points out the redirection attacks, man-in-the-middle attacks, and endpoint attacks these systems may face.
Rahman et al.(2020)~\cite{rahman_adversarial_2020}discusses the risks of adversarial example attacks on COVID-19 deep learning systems and notes that such attacks could impact the security of medical Internet of Things devices.
Gongye et al.(2020)~\cite{gongye_new_2020} identifies passive and active attacks on deep neural networks in medical applications, including adversarial training attacks, data replay attacks, and model extraction attacks, and conducts empirical validation.
Tag et al.(2023)~\cite{tag_ddod_2023} studies the adversarial decision blocking attacks human-AI teams may face in medical decision-making scenarios and indicates these attacks could influence AI-assisted medical judgments.

Additionally, the transparency and interpretability issues of AI systems may make it difficult for doctors and patients to understand and trust the decision-making process of these systems. From a legal and ethical perspective, the application of AI technology in the medical field also faces risks, including the unclear legal status of medical AI, disputes over liability attribution, and ethical challenges related to patient privacy protection, medical safety, and responsibility allocation.

\subsubsection{Political scenarios}
Several studies have explored the role of AI in shaping electoral politics and democratic processes. Yu (2024) ~\cite{yu_how_2024}examined how AI could potentially steal elections through manipulation of social media and voting systems. Klein (2024)  ~\cite{klein_impact_nodate} discussed the impact of AI-driven social media on voter autonomy and government. Schippers(2020)  ~\cite{schippers_artificial_2020} provided an overview of the implications of artificial intelligence for democratic politics. Kamal et al.(2024)~\cite{ kamal_artificial_2024} investigated how AI-powered political advertising could harness data-driven insights for campaign strategies. Filgueiras (2022)  ~\cite{filgueiras_politics_2022} analyzed the politics of AI in developing countries in relation to democracy and authoritarianism. Tomar et al.(2024)~\cite{tomar_role_nodate}  studied the role of AI tools in influencing Indian elections and social media dynamics. Jungherr (2023)~\cite{jungherr_artificial_2023} proposed a conceptual framework for understanding the relationship between artificial intelligence and democracy. Starke and Lünich (2020)  ~\cite{starke_artificial_2020} analyzed the effects of AI on citizens’ perceptions of political legitimacy in the EU. Several papers also discussed broader topics such as the transformative role of AI in reshaping electoral politics~\cite{munoz_transformative_2023}  and the impact of financial crises on elections ~\cite{magalhaes_introduction_2014}. In summary, this body of research has begun to uncover both opportunities and risks associated with the growing role of advanced technologies in modern political systems and campaigning.

The research focus of SE attacks in political contexts has evolved from technical to policy aspects, with more attention paid to how AI may impact democratic political structures and citizens' participation levels. Emerging research directions primarily concern the manipulation of public opinion environments and voter preferences through social media influence operations, and applications in developing countries and different regimes, exploring the distinct socio-political impacts of technologies under democratic and authoritarian systems. Future research needs to examine how AI-driven SE attacks may influence the evolution of political ecosystems.
\section{Challenges and prospects}\label{sec4}

Based on our comprehensive review of the evolution of AI-powered SE attack methods through the 3E phases(enlargement, enrichment and emergence), as well as the analysis of research trends and gaps discussed in prior sections, we now explore current challenges and future perspectives. These emerging research avenues provide a deeper understanding of complex issues in SE defense while addressing recent challenges and promoting comprehensive, in-depth progress. Building on prior analyses, we aim to facilitate deeper understanding of evolving AI risks and corresponding defense strategies. 

\subsection{Challenges}

As AI and its potential misuse rapidly evolve, the research community faces several key challenges in addressing growing threats from AI-powered SE attacks. Overcoming these requires multifaceted, proactive efforts ensuring AI system and application security and resilience.

The first challenge concerns tracking and analyzing the rapidly changing AI-enabled SE attack landscape. Given AI's ongoing evolution, maintaining awareness of latest developments and impacts on social engineering poses difficulty.
Second, developing a robust framework encompassing LLM's transformative role in SE attacks remains pertinent. While the work by Alahmed et al. (2024)
~\cite{alahmed2024exploring} provides a conceptual model focused on LLM, there is a need to explore other aspects of AI that can be misused in SE attacks, beyond just content generation.
Third, proactive, adaptive detection and defense strategies must pace the evolving SE attack threatscape. With the rapid evolution of AI-enabled SE attacks, there is a need to develop detection and mitigation strategies that are proactive, adaptive, and can keep pace with the changing threat landscape.
Finally, ethical and privacy implications raised through AI-enabled SE attacks necessitate careful consideration within research agendas. Unaddressed, such concerns endanger populations and trust in AI progress.

\subsection{Future direction}

\subsubsection{Quantitative risk assessment of AI-powered SE attacks}

With advances in AI technologies, it has become imperative to rigorously evaluate their broader impacts on society through augmented social engineering  attacks. A direction for future research is developing a standardized risk quantification framework for AI-empowered SE attacks.
Such a framework could serve as a foundation for: 1) conducting comparative impact assessments of how AI shapes emerging forms and trends of SE attacks under different sociotechnical settings and application scenarios, 2) enabling forensic traceability of factors that exacerbate social risks, and 3) prioritizing mitigation measures in a principled manner. 

Previous work in related domains provides a starting point. For instance, Markov decision process frameworks have been applied to autonomous vehicle safety verification~\cite{hejase2020methodology}. Causal Bayesian networks have also been used to establish risk profiles for AI systems and analyze different risk dimensions~\cite{khan2023efficient}. Moreover, state space models have formed the basis of frameworks for assessing AI system robustness and explainability~\cite{novelli2023evaluate}.

This study aims to develop a standardized risk quantification framework based on Markov decision process (MDP) for AI-powered SE attacks as a foundation for comparative impact evaluations, forensic traceability of sociotechnical factors, and prioritization of mitigation measures.
The framework could augment SE attack propagation dynamics and causally influence downstream social impacts over time. This involves leveraging big data to track changes in key risk metrics like spreading capability and penetration efficiency under different conditions.

\paragraph{Problem Definition}
The SE attack in social system can be modeled as a Markov decision process. For each attack action, the policy $\pi_t^i$ is made based on the current system state $s_t^i$ and observation $o_t^i$, in order to maximize the long-term reward or satisfy a specific objective. These decisions include the attack actions. Additionally, the social network relationships in the system also influence the risk state transitions $P(s'|s,a)$ and the reward structure $R^i(s,a)$, resulting in a complex dynamic system.

For the social system, the Markov decision process can be represented as $\mathcal{MDP} = (\mathcal{S}, \mathcal{A}, \mathcal{P}, \mathcal{R}, \mathcal{O}, \Gamma, \Pi)$, where:
\begin{itemize}
    \item  $\mathcal{S}$ is the set of state space, which represents the possible configurations of the social system.
    \item $\mathcal{A}$ is the set of action space. The attack action $a$ can be a specific social engineering technique, such as sending phishing emails or creating fake websites. 
    \item $\mathcal{P}$ represents the state transition probability, which captures the dynamics of the social system. Given the set of target states $S_t = \{s_t^1, s_t^2, \ldots, s_t^N\}$ and the set of actions $A_t = \{a_t^1, a_t^2, \ldots, a_t^N\}$, the probability of the target $i$ transitioning from $s_t^i$ to $s_{t+1}^i$ is $P(s_{t+1}^i|s_t^i,a_t^i)$. The state transition probability for the entire system is:
    \begin{equation}
    P(S_{t+1}|S_t, A_t) = \prod_{i \in \mathcal{T}} P(s_{t+1}^i | s_{t}^i, a_t^i)
    \label{eq:state_transition}
    \end{equation}
    \item $\mathcal{R}:\mathcal{S} \times \mathcal{A} \rightarrow \mathbb{R}$ is the reward function, which maps state-action pairs to real-valued rewards received by the attacker. This models the objectives of the SE attack.
    \item $\mathcal{O}$ is the observation space set, denoting the possible observations the attacker can receive about the current state of the social system.
    \item $\Gamma$ is the discount factor, which represents the discounting rate of future rewards.
    \item $\Pi$ the set of possible policies, which are mappings from states and observations to attack actions. 
\end{itemize}
\paragraph{Metrics for measuring the risk of SE attacks}

\textbf{Spreading capability}
Let us define the random variable $\mathbf{S} = (S^1, S^2, \ldots, S^N)$ to represent the system state, where $S^i$ denotes the infection state of user $i$, and its value can range from $0$ to $d_{\max}$.

The system state distribution is defined as $P(\mathbf{S} = \mathbf{s}) = P(S^1 = s^1, S^2 = s^2, \ldots, S^N = s^N)$, which represents the probability of the system being in state $\mathbf{s} = (s^1, s^2, \ldots, s^N)$.

The entropy of the system state is defined as:

\begin{equation}
H(\mathbf{S}) = -\sum_{\mathbf{s}} P(\mathbf{S} = \mathbf{s}) \log P(\mathbf{S} = \mathbf{s})
\end{equation}

The entropy reflects the uncertainty of the system state, and a higher entropy indicates a more uniform distribution of the system states, suggesting a broader spread of the attack propagation.

Furthermore, we can define the Kullback-Leibler (KL) divergence to quantify the change in the system state distribution:

\begin{equation}
D_{\mathrm{KL}}(P_1 | P_0) = \sum_{\mathbf{s}} P_1(\mathbf{s}) \log \frac{P_1(\mathbf{s})}{P_0(\mathbf{s})}
\end{equation}

where $P_0$ and $P_1$ represent the system state distributions before and after the attack, respectively. 
\begin{equation}
\beta=D_{\mathrm{KL}}(P_1 | P_0) - D_B 
\end{equation}

where $D_B$ is the baseline referring to the natural change of state distribution over time if no attack occurs. A larger spreading capability $\beta$ indicates a more significant change in the system state distribution due to the attack.







\textbf{Penetration efficiency}
The penetration efficiency $\eta$ represents the ability of the attacker to maximize the impact of the attack per unit of their own resources~\cite{ghanem2018reinforcement}. It can be defined as the ratio of the expected reward obtained from a successful attack to the cost incurred by the attacker:



\begin{equation}
\eta = \frac{\mathbb{E}[R(s_{t+1}, a_t^i) | s_t^i, a_t^i]}{\mathbb{E}[C(a_t^i) | s_t^i, a_t^i]}
\end{equation}

where $R(s_{t+1}, a_t^i)$ is the reward function that represents the benefit obtained from the attack, and $C(a_t^i)$ is the cost function that represents the resources consumed by the attacker.

The numerator, $\mathbb{E}[R(s_{t+1}, a_t^i) | s_t^i , a_t^i]$, represents the expected reward obtained from a successful attack, given that the attacker  takes the action $a_t^i$. The denominator, $\mathbb{E}[C(a_t^i) | s_t^i , a_t^i]$, represents the expected cost incurred by the attacker for taking the action $a_t^i$ .

The penetration efficiency $\eta$ is a measure of the attacker's ability to maximize the impact of the attack per unit of their own resources. A higher penetration efficiency indicates that the attacker can achieve a greater impact with fewer resources, which can be an important factor in the evaluation and optimization of attack strategies.

These indicators reflect the capability level of social engineering attacks from different dimensions. Defining these metrics helps to
quantitatively assess the risk levels of different attack methods, detect weak points in each link of the attack chain and design targeted defenses and
evaluate the effectiveness of defense strategies and prioritize improvements to security capabilities.

\subsubsection{Defense techniques against AI-powered SE attack}

Addressing AI-empowered social engineering attacks will benefit from rigorous investigation into robust, multidimensional defensive strategies. Promising research directions involve developing detection systems capable of identifying and mitigating malicious AI-generated content across various networks and platforms.
Research can focus on developing efficient AI detection algorithms and establishing response mechanisms to cope with evolving SE attack threats~\cite{parthy_identification_2019,syafitri_social_2022}.
Table \ref{tab:defense} outlines the technical measures to defend against SE attack risks, categorized from the perspectives of the attack targets and the AI life cycle stages. This table provides a comprehensive overview of the technical measures that can be employed to defend against SE attack risks from multiple perspectives.

\paragraph{Attack targets}
At the individual level, anonymization techniques like data perturbation and encryption can help protect privacy from personal data aggregation and profiling~\cite{xue_advparams_2022}, as summarized in Table \ref{tab:defense}. Personalized monitoring tools that identify anomalous behaviors may also help compensate for technical limitations~\cite{turgay_perturbation_2023}. At the organization level, computational models can analyze sentiment trends and flag deceptive narratives spread through social media. Content filtering algorithms can timely recognize and address misinformation propagation~\cite{mazurczyk_disinformation_2024,jain_online_2021}. Cross-validation of information from multiple sources through aggregation and fusion supports coordinated responses across groups~\cite{shrivastava_defensive_2020}.  
At the community level, recommendation diversification and alternative information platforms aim to prevent echo chambers and support balanced views~\cite{hu_optimal_2020}.
Participatory digital governance empowers communities for self-organized defense~\cite{kurniawan_attck-kg_2021}. Continued advancement across these technical defensive layers is needed to reinforce system integrity and user trust against increasingly sophisticated AI-driven attacks.

\paragraph{Risk defense across the AI lifecycle}

Comprehensively defending against AI-powered SE attacks requires constructing layered security that spans the full AI development lifecycle.
Defense techniques against known threats include adversarial training, network distillation, adversarial sample detection, deep neural network(DNN) verification, data filtering, integrated analysis, pruning, and differentially-private teacher models. Federated learning updates models while protecting privacy and harnessing distributed computation against centralization abuse.
Another important direction is adversarial training, as demonstrated by the ALUM method proposed in ~\cite{liu_adversarial_2020}. ALUM applies perturbations in the embedding space to regularize the training objective, achieving gains in both generalization and robustness for large language models.
Staliunaite et al.(2021)~\cite{staliunaite_improving_2021} showed that combining adversarial training and data augmentation can enhance the performance of commonsense causal reasoning models.
Knowledge distillation has emerged as an effective approach for model compression and transfer)~\cite{song_lightpaff_2020}. 
Going further, Padmanabhan et al. (2024)~\cite{padmanabhan_propagating_2024} presented a distillation-based method that not only injects entity knowledge but also propagates it to enable broader inferences.
Differential privacy has been recognized as a crucial technique for preserving the privacy of training data in NLP models~\cite{flemings_differentially_2024}. 
The survey ~\cite{hu_differentially_2023} further discusses the recent advances and future directions of differentially private NLP models.

Model security faces adversarially transferred vulnerabilities requiring enhanced detectability, verifiability and explainability against unconceived threats. 
In terms of detectability, researchers have proposed techniques such as watermarking to embed detectable signals in model outputs, which can help identify their use for SE attacks, e.g., detecting maliciously generated content like disinformation or explicit material, to prevent model misuse. On verifiability, traceable information flows verify content accuracy by external corroboration, enhancing trustworthiness. Improving model explainability can also enhance user understanding of model behavior and outputs, increasing trust and acceptance. Better diagnosis and correction of model errors or biases can also enhance model security.

In the area of detectability, the HuntGPT system ~\cite{ali_huntgpt_2023} utilizes a random forest classifier and an XAI ~\cite{zytek_llms_2024}framework to detect and explain network anomalies using large language models (LLMs). Additionally, the "Data-to-Paper" automated platform ~\cite{ifargan_autonomous_2024} provides programmable traceability, and the "Quote-Tuning" method ~\cite{zhang_verifiable_2024} employs LLMs to verify generated content through verbatim referencing. For explainability, the integration of LLMs with decision models ~\cite{goossens_integrating_2023}and the translation of AI algorithm explanations into natural language  can improve the interpretability and usability of AI. Furthermore, research on LLM watermarking has proposed the "WaterMax" scheme ~\cite{giboulot_watermax_2024} and the open-source "MarkLLM" toolkit~\cite{pan_markllm_2024} to balance watermark detectability, robustness, and generation quality.



Regarding architectural security, it is crucial to defense the potential security risks introduced by AI systems by comprehensively utilizing isolation, detection, circuit breaking, and redundancy mechanisms to enhance the robustness of business products.
The emerging paradigm of structured interactions limits the way users interact with LLM systems~\cite{shevlane_structured_2022}, preventing the widespread use of dangerous functionalities while preserving secure utilization. SecGPT~\cite{wu_secgpt_2024}is an execution isolation architecture that isolates the execution of LLM-based applications and precisely mediates their interactions outside the isolated environment, mitigating security and privacy risks. ShieldLM~\cite{zhang_shieldlm_2024} utilizes LLM as an aligned, customizable, and interpretable security detector to enhance the security and reliability of LLM systems. The deployment correction ~\cite{obrien_deployment_2023} proposes an incident response framework for identifying and rectifying potential issues that may arise during the deployment of advanced AI models. The safety case~\cite{clymer_safety_2024} provides a structured approach to demonstrating the safety of advanced AI systems, considering arguments such as the inability to cause catastrophe, robust control measures, and trustworthiness. The critical infrastructure protection ~\cite{yigit_critical_2024}explores methods to leverage LLM to enhance the security and resilience of critical national infrastructure, addressing challenges related to trust, privacy, and security.

\begin{table}[htbp]
\centering
\caption{SE attack defense method.}
\label{tab:defense}
\begin{tabular}{p{1cm}p{1.5cm}p{4cm}p{1cm}}

\toprule
Defense & Categories & Technique & Reference
\\ 
\hline
\multirow{6}{1cm}{Attack targets} & Individuals & Data perturbation and encryption & ~\cite{xue_advparams_2022}
\\ 
& \multirow{3}{2cm}{Organizational Entities}  & Computational   propaganda models & ~\cite{mazurczyk_disinformation_2024}
\\ 
&  & Content detection & ~\cite{jain_online_2021}
\\ 
&  & Multi-source data aggregation and fusion  & ~\cite{shrivastava_defensive_2020}
\\ 
& \multirow{2}{2cm}{Community Groups}   & Diverse information   platforms and recommendation & ~\cite{hu_optimal_2020}
\\ 
&  & Digital civic participation and governance  & ~\cite{kurniawan_attck-kg_2021}
\\ 
\hline
\multirow{12}{1cm}{Risk defense across the AI lifecycle} & \multirow{5}{2cm}{Attack-defense techniques} & Adversarial training   & ~\cite{liu_adversarial_2020}
\\ 
 &  & Data augmentation & ~\cite{staliunaite_improving_2021}
 \\ 
 &  & Knowledge distillation  & ~\cite{song_lightpaff_2020}
 \\ 
 &  & Distillation-based method  & ~\cite{padmanabhan_propagating_2024} 
 \\ 
 &  & Differential privacy  & ~\cite{flemings_differentially_2024} 
 \\ 
 & \multirow{3}{2cm}{Model security}  & Detectability-watermarking &  ~\cite{giboulot_watermax_2024}
 \\ 
 &  & Verifiabilitythe -"Quote-Tuning" method  & ~\cite{zhang_verifiable_2024}
 \\ 
 &   & Explainability & ~\cite{zytek_llms_2024}
 \\ 
& \multirow{4}{2cm}{Architectural security} & Structured   interactions & ~\cite{shevlane_structured_2022}
\\
 &  & Isolation architecture  & ~\cite{wu_secgpt_2024}
 \\ 
&  & Detection  & ~\cite{zhang_shieldlm_2024}
\\ 
 & & Circuit-breaking and redundancy & ~\cite{obrien_deployment_2023} 
 \\ 
\bottomrule
\end{tabular}
\end{table}

\subsubsection{Ethical and legal efforts in AI application environments}

Given the potential for societal harms from AI-enabled social engineering attacks, further work could examine associated ethical and legal considerations. Research is needed to establish frameworks for responsible AI integration into networked human systems and accountability of developers. Normative constraints on AI-facilitated social manipulation should also be explored within legal and regulatory contexts.

Clear allocation of responsibilities across AI system lifecycles from data curation to deployment is imperative for effective governance. Simulation-based modeling of complete development pipelines allows assessing accountability at each stage and comparing effects of oversight approaches. Such computational analyses could evaluate mitigation strategies for AI-driven social threats under different legislative regimes.

Continued methodological advancement is required, such as algorithmic auditing, impact assessments within virtual environments, and mechanisms for establishing AI's legal duties and accountability for potential harms. Perception modules in simulators should identify adverse societal outcomes to guide responsible innovation. Considering AI augmentation from enforcement perspectives regarding defined legal status facilitates addressing issues of redress for threats posed by emergent attack vectors.
Ongoing multidisciplinary work is needed to embed ethical safeguards and accountability as the interface between human and artificial social actors grows in scope and complexity.
\section{Conclusion}\label{sec5}

The study delineated the 3E progression of AI-powered social engineering attacks, encompassing Enlarging reach via digitization, Enriching tailored vectors, and the possible Emergence of innovative deception modes employing LLM.
Existing SE approaches were classified into technological eras, with representative cases discussed to illustrate impacts. 

Several key challenges in the field were identified, including the difficulty of systematically tracking shifting attack paradigms, developing a robust framework encompassing AI risk, and ensuring defenses remain  adaptive to attack methods. A Markov decision process based framework was proposed to facilitate quantitative risk assessment.
To address limitations in prior work, this study consolidated extant technical investigations while proposing framework to systematically address quantitative risk assessment. Specifically, we compared underlying exploitation methods, outlined implementation characteristics, and analyzed threat progression.

Future directions were also highlighted emphasizing the importance of quantitative risk assessment of AI-powered SE attacks, developing defense techniques against AI-powered SE attack and efforts spanning both technical and policy domains. Multifaceted research is required to curb foreseeable exploits preemptively through vigilant monitoring, evidence-based policymaking, and comprehensive stakeholder engagement.

We hope our thorough investigation into the evolving landscape of social engineering attacks in the context of artificial intelligence, offering valuable insights and charting future research directions.

\newpage
\balance

\printbibliography

\newpage
\onecolumn

\end{document}